\documentclass[12pt]{article}
\usepackage{color}
\usepackage{amssymb}
\usepackage{graphicx}
\usepackage{epstopdf}
\usepackage[hang,small,bf]{caption}
\usepackage{hyperref}
\usepackage{amsmath}
\usepackage{latexsym}
\usepackage{setspace}

\newcommand{\be}[3]{\begin{equation}  \label{#1#2#3}}
\newcommand{\ee}{\end{equation}}
\newcommand{\ba}{\begin{array}}
\newcommand{\ea}{\end{array}}
\newcommand{\bea}[3]{\begin{eqnarray}  \label{#1#2#3}}
\newcommand{\eea}{\end{eqnarray}}

\let\Large=\large
\let\large=\normalsize

\newcommand{\haken}{\mathbin{\hbox to 8pt{%
                 \vrule height0.4pt width7pt depth0pt
                 \kern-.4pt
                 \vrule height4pt width0.4pt depth0pt\hss}}}

\renewcommand{\theequation}{\thesection.\arabic{equation}}

\def\openone{\leavevmode\hbox{\small1\kern-3.8pt\normalsize1}}

\def\bo{{\raise.15ex\hbox{\large$\Box$}}}               
\def\face{{\raise.2ex\hbox{$\displaystyle \bigodot$}\mskip-2.2mu \llap {$\ddot
        \smile$}}}                                      

\def\leftrightarrowfill{$\mathsurround=0pt \mathord\leftarrow \mkern-6mu
        \cleaders\hbox{$\mkern-2mu \mathord- \mkern-2mu$}\hfill
        \mkern-6mu \mathord\rightarrow$}       
\def\dvec#1{\vbox{\ialign{##\crcr
        \leftrightarrowfill\crcr\noalign{\kern-1pt\nointerlineskip}
        $\hfil\displaystyle{#1}\hfil$\crcr}}}           

\def\beq{\begin{equation}}
\def\eeq{\end{equation}}

\def\beqx{\begin{displaymath}}
\def\eeqx{\end{displaymath}}

\def\beqa{\begin{eqnarray}}
\def\eeqa{\end{eqnarray}}

\begin{document}
\DeclareGraphicsExtensions{.jpg,.pdf,.mps,.eps,.png}
\begin{flushright}
\baselineskip=12pt
EFI-09-30 \\
ANL-HEP-PR-09-102
\end{flushright}

\begin{center}
\vglue 1.5cm

{\Large\bf Prospects for Higgs Searches at the Tevatron and LHC in the MSSM with Explicit $CP$-violation } \vglue 2.0cm {\Large Patrick Draper$^{a}$, Tao Liu$^{a}$, and Carlos E.M. Wagner$^{a,b,c}$}
\vglue 1cm {
$^a$ Enrico Fermi Institute and
$^b$ Kavli Institute for Cosmological Physics, \\
University of Chicago, 5640 S. Ellis Ave., Chicago, IL
60637\\\vglue 0.2cm
$^c$ HEP Division, Argonne National Laboratory,
9700 Cass Ave., Argonne, IL 60439
}
\end{center}

\vglue 1.0cm
\begin{abstract}
We analyze the Tevatron and Large Hadron Collider (LHC) reach for the Higgs sector
of the Minimal Supersymmetric Standard Model (MSSM) in the
presence of explicit $CP$-violation. Using the most recent studies
from the Tevatron and LHC collaborations, we examine the CPX
benchmark scenario for a range of $CP$-violating phases in the soft
trilinear and gluino mass terms and compute the exclusion/discovery
potentials for each collider on the $(M_{H^+}, \tan\beta)$ plane.
Projected results from Standard Model (SM)-like, non-standard, and
charged Higgs searches are combined to maximize the statistical
significance. We exhibit complementarity between the SM-like Higgs searches at the LHC with low luminosity and the Tevatron, and estimate the combined reach of the two colliders in the early phase of LHC running.
\end{abstract}

\newpage

\section{Introduction}

The origin of electroweak symmetry breaking remains a principal open question in high-energy physics. In the Standard Model (SM), the breakdown of the electroweak symmetry is induced by the vacuum
expectation value of a scalar field, which transforms non-trivially
under the $SU(2)_L\times U(1)_Y$ symmetry. A consequence of this
mechanism of spontaneous symmetry breaking is the presence of a physical
scalar Higgs particle, with well-defined couplings to fermions and
gauge bosons. The search for such a particle at lepton and hadron
colliders is therefore a paramount goal in particle
physics. The LEP experiments have already excluded at 95\% C.L. the presence of a SM-like
Higgs with mass below 114.4~GeV~\cite{Barate:2003sz}.
In the coming years, Higgs searches will be performed at
hadron colliders.

The Tevatron collider at Fermilab has an active Higgs search program and
has already excluded a SM-like Higgs at 95\% C.L. in the mass range 160--170~GeV~\cite{:2009pt}.
The Tevatron is expected to operate until the end of 2011. It is likely that by this time
the CDF and D0 experiments will collect 10--12~$\mbox{ fb}^{-1}$
apiece and achieve
some improvements in the analysis, yielding a significant chance that they
will be able to probe the entire SM Higgs mass range
$110-190\mbox{ GeV}$. A similar conclusion is reached for the parameter
space of the Higgs sector in the $CP$-conserving MSSM, provided that
the limits derived from SM-like Higgs searches are statistically combined
with limits from direct searches for the non-standard Higgs
bosons~\cite{Draper:2009fh}.

The Large Hadron Collider (LHC) at CERN will begin collisions at
the end of 2009
with an expected center-of-mass energy of several TeV, and the ATLAS and
CMS experiments will collect on the order
of a few hundred pb$^{-1}$ of data during 2010.
A higher center of mass energy of $14\mbox{ TeV}$ is expected to be
achieved after this run, once the necessary upgrades are completed.
The anticipated rate of data acquisition at $14 \mbox{ TeV}$ in the early years of the LHC is expected to be
a few to 10~fb$^{-1}$/year. Once the LHC acquires a few fb$^{-1}$ of data
at $14\mbox{ TeV}$, the Tevatron
and LHC reaches for a light Higgs boson may be comparable and even
complementary in some searches.
Eventually, once
the LHC experiments collect 10 to 30~fb$^{-1}$ of $14\mbox{ TeV}$ data,
the LHC will probe the SM and MSSM Higgs sectors at high statistical
significance, far superior to what is attainable at the Tevatron~\cite{Aad:2009wy,Ball:2007zza}.

In this study, we analyze the Tevatron and LHC reach for the MSSM Higgs
sector~\cite{ERZ}--\cite{mhiggsFD2} in the presence of explicit $CP$-violation~\cite{Pilaftsis:1998dd}--\cite{Ibrahim:2002zk}.
At the Tevatron we consider 10 fb$^{-1}$ in all channels
and provide projections for a set of
possible improvements in signal efficiencies. For the LHC, to study
the possible initial complementarity with the Tevatron results, we
consider the case of 3 fb$^{-1}$ at $14\mbox{ TeV}$. To display the
long-term LHC
capabilities, we also show the results for
30 fb$^{-1}$ at $14\mbox{ TeV}$ in the most challenging scenario.
Let us mention that
there are alternative
possibilities for the LHC timeline where the collision energy is kept below $14\mbox{ TeV}$ for several years, and the upgrade to $14\mbox{ TeV}$ is
only completed later. We do not attempt an analysis of these scenarios
because we base our study on the
Higgs reach projections presented by the LHC collaborations, which
are only fully complete for a center of mass energy of $14\mbox{ TeV}$.
For reference, preliminary results indicate that to obtain the same Higgs reach, the
luminosity at $10\mbox{ TeV}$ should be twice as large as that
which is required at 14~TeV~\cite{Delmastro:2009wc}.

Our work differs in three significant ways from previous analyses of
Higgs searches in the MSSM with
explicit $CP$-violation
performed in
Refs.~\cite{Carena:2002bb,Accomando:2006ga}.
First, for the Tevatron we use the 2009 limits from the CDF
and D0 experiments given in
Refs.~\cite{CDFSM}--\cite{cdfcharged}. For the LHC
we incorporate the projections presented by the experimental collaborations
in the most recent technical design and expected physics performance
documents~\cite{Aad:2009wy,Ball:2007zza}. These projections show
marked differences from the earlier TDRs, and as a result the priority
for some channels has been reduced, while others have been elevated.
Secondly, for both colliders we analyze the potential for the non-standard MSSM Higgs searches, and provide the combination with the SM-like Higgs reach. These two types of searches offer
considerable complementarity and together can be used to cover most of the
analyzed parameter space. For the Tevatron, we also include the reach for the charged Higgs. Finally, we present the combination of the Tevatron discovery reach with the LHC reach at 3~fb$^{-1}$. At this low LHC integrated luminosity, the statistical significances offered by the two colliders may be comparable, and so it may be of interest to perform the combination. We present this analysis only for the SM-like Higgs search channels, which offer greater complementarity than the non-standard channels.

For the LHC reach we compute and combine discovery significances using Poisson statistics and the profile likelihood ratio, evaluated on data fixed to the expected values for the signals and backgrounds. For the Tevatron we work with 95\% C.L. upper bounds on the signal presented by CDF and D0, combining them in inverse quadrature. This combination method is strictly valid only in the Gaussian limit; however, it was tested in Ref.~\cite{Draper:2009fh} and found to match well with the combination derived from a full analysis performed by the collaborations. A further discussion of statistical methods is presented in Appendix A.

In the MSSM Higgs sector, $CP$-violation can occur via the incorporation of explicit phases in the supersymmetry breaking parameters.
$CP$-violating phases can be removed from the tree-level Higgs
potential by field redefinitions. However, phases in the soft trilinear couplings and the gaugino mass
terms influence the effective Higgs Lagrangian through loop corrections~\cite{Pilaftsis:1998pe,Pilaftsis:1999qt}.
We shall work in the CPX benchmark scenario, defined by the following
parameter values at the soft scale~\cite{Carena:2000yi}:
\begin{eqnarray*}
M_S=500\mbox{ GeV, }& &|A_t|=1\mbox{ TeV,}\nonumber\\
\mu=2\mbox{ TeV, }& &M_{1,2}=200\mbox{ GeV,}\nonumber\\
A_{b,\tau}=A_t\mbox{, }& &|M_{\tilde{g}}|=1\mbox{ TeV}.
\end{eqnarray*}
In the above, $A_f$ are the trilinear Higgs sfermion couplings,
$M_S$ is the characteristic scale of soft supersymmetry breaking
scalar masses, and $M_{\tilde{g}}$ is the gluino mass.
We set the top quark mass to $m_t=173.1\mbox{ GeV}$. We shall perform our
analysis scanning over the charged Higgs
mass $M_{H^+}$ and $\tan\beta$ over the ranges
$(100\mbox{ GeV},400\mbox{ GeV})$ and $(2,60)$, respectively, for a
variety of complex phases of $A_{t,b,\tau}$ and $M_{\tilde{g}}$\footnote{If we began with phases for these parameters at a higher scale, phases for first and second generation trilinear parameters would be generated by RG running, which are highly constrained by EDM measurements~\cite{Garisto:1996dj}. This can be avoided either by fixing these phases to zero at the weak scale, or by increasing the soft masses of the first and second generation sfermions. However, note that the first and second generation parameters do not have a significant influence on the results presented in this work.}. Masses, mixings, and branching ratios are computed with CPsuperH~\cite{Lee:2007gn} and HDECAY~\cite{Djouadi:1997yw}; SM Higgs cross sections are taken from Ref.~\cite{maltoni} and rescaled to obtain cross sections in the MSSM. The dominant effects of the phases are twofold. First, they cause the neutral Higgs
mass eigenstates to become admixtures of $CP$-even and $CP$-odd components, modifying
the couplings to gauge bosons relative to those in the case with no
$CP$-violation. For example, the lightest neutral Higgs can now have a
significant $CP$-odd component, strongly suppressing its couplings to the
$W$ and $Z$ bosons. Secondly, the Yukawa couplings are altered, leading
in particular to modifications of the neutral Higgs decay branching ratios to $b\bar{b}$ and
$\tau^+\tau^-$, and different production cross sections through the
bottom quark fusion and gluon fusion mechanisms. We examine separately
the reach in those channels designed to search for a Higgs with SM-like
gauge couplings (hereafter referred to as an ``SM-like Higgs"), and in those
channels which probe either neutral scalars with negligible gauge couplings
(hereafter, ``non-standard Higgs") or charged Higgs states, in order to understand the
complementarity of their coverages. Afterwards we combine the statistical significances of all channels to obtain an overall reach in the MSSM Higgs parameter space.

Our presentation is organized as follows. In section 2 we review the
couplings of the MSSM effective Lagrangian that are of particular
relevance for understanding the CPX reach. In section 3 we present
and analyze the results for the Tevatron and make conservative estimates
of the improvements in signal efficiency necessary to cover large regions
of parameter space. Section 4 contains the projections for the LHC, and we
offer conclusions in section 5. Appendix A offers a brief review of statistical methods, followed by a discussion of the
approximations used in the text to compute and combine exclusion limits
and discovery significances for multiple channels at the Tevatron and LHC. Appendix B extends some of the discussion in the text to the case of $CP$-conserving benchmark scenarios.

\section{Effective Yukawa Couplings}

The radiative corrections to the Yukawa couplings of Higgs states to
down-type fermions~\cite{Carena:1998gk}--\cite{Carena:2001bg} play a significant role in inducing $CP$-violating
effects in the Higgs sector and can strongly affect the SM-like Higgs search
channels at colliders.
The scalar and pseudoscalar neutral Higgs couplings to bottom quarks in the
effective Lagrangian are given by~\cite{Carena:2002bb}
\begin{equation}
\mathcal{L}=-g_fH_i\bar{b}(g^S_{H_ib\bar{b}}+\imath\gamma_5g^P_{H_ib\bar{b}})b
\end{equation}
where $g_f$ is the SM scalar coupling given by the bottom quark mass over the vacuum expectation value of the Higgs, and $g^{S,P}$ are given by
\begin{eqnarray}
g^S_{H_ib\bar{b}}&=&\mbox{Re}\left(\frac{1}{1+\kappa_b\tan\beta}\right)
\frac{\mathcal{O}_{1i}}{\cos\beta}+\mbox{Re}
\left(\frac{\kappa_b}{1+\kappa_b\tan\beta}\right)
\frac{\mathcal{O}_{2i}}{\cos\beta} \nonumber\\
&+&\mbox{Im}\left(\frac{\kappa_b(\tan^2\beta+1)}{1+\kappa_b\tan\beta}\right)
\mathcal{O}_{3i}\nonumber\\
g^P_{H_ib\bar{b}}&=&-\mbox{Re}\left(\frac{\tan\beta-\kappa_b}
{1+\kappa_b\tan\beta}\right)\mathcal{O}_{3i}
+\mbox{Im}\left(\frac{\kappa_b\tan\beta}{1+\kappa_b\tan\beta}\right)
\frac{\mathcal{O}_{1i}}{\cos\beta} \nonumber\\
&-&\mbox{Im}\left(\frac{\kappa_b}{1+\kappa_b\tan\beta}\right)
\frac{\mathcal{O}_{2i}}{\cos\beta}.
\label{effcoup}
\end{eqnarray}
Here $\mathcal{O}_{jk}$ is the neutral Higgs mixing matrix, where $j$
is associated with the gauge eigenstate $\{H^0_u,H^0_d,A\}$ and
$k$ runs over the
mass states $\{H_1,H_2,H_3\}$ which are ordered so that $(M_{H_3}\geq M_{H_2}\geq M_{H_1})$, and $\kappa_b$ parameterizes the
radiative contributions from sbottom-gluino and stop-chargino loops,
\begin{eqnarray}
\kappa_b&=&\frac{(\Delta h_b/h_b)}{1+(\delta h_b/h_b)}\nonumber\\
\Delta h_b/h_b&=&\frac{2\alpha_s}{3\pi}
M_{\tilde{g}}^*\mu^*I(m^2_{\tilde{d}_1},m^2_{\tilde{d}_2},|M_{\tilde{g}}|^2)
+\frac{|h_u|^2}{16\pi^2}A_u^*\mu^*
I(m^2_{\tilde{u}_1},m^2_{\tilde{u}_2},|M_{\tilde{g}}|^2)\nonumber\\
\delta h_b/h_b&=&-\frac{2\alpha_s}{3\pi}M_{\tilde{g}}^*A_b
I(m^2_{\tilde{d}_1},m^2_{\tilde{d}_2},|M_{\tilde{g}}|^2)
-\frac{|h_u|^2}{16\pi^2}|\mu|^2
I(m^2_{\tilde{u}_1},m^2_{\tilde{u}_2},|M_{\tilde{g}}|^2)
\end{eqnarray}
where $I(a,b,c)$ is a function that behaves as $1/\max(a^2,b^2,c^2)$~\cite{deltamb1}--\cite{deltamb2b}.
The size of the dominant loop corrections to the Higgs sector is controlled
by $\mu$, so a large value of $|\mu|$ is taken in CPX to accentuate the
$CP$-violating effects.

\begin{figure}[!htbp]
\begin{center}
\begin{tabular}{cc}
\includegraphics[width=0.45\textwidth]{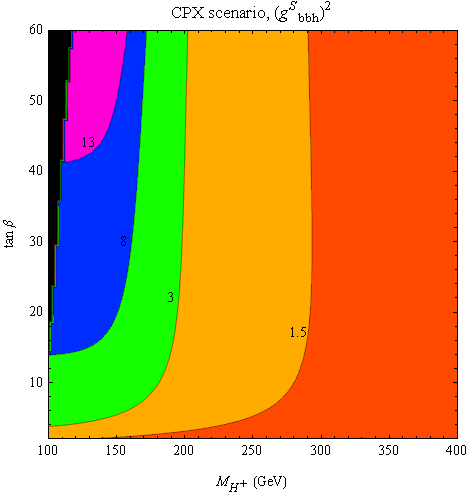} &
\includegraphics[width=0.45\textwidth]{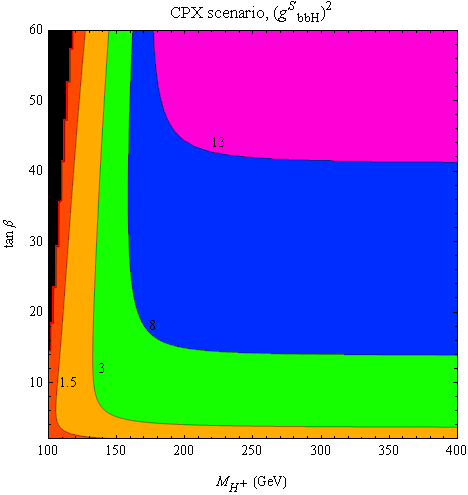}
\end{tabular}
\caption{Effective bottom Yukawa couplings squared for the $CP$-even Higgs states $h$ and $H$ in the $CP$-conserving
limit of CPX.}
\label{effhbb}
\end{center}
\end{figure}

For illustration, let us consider the behavior of the effective couplings
in the simplest case of vanishing phases. In this scenario $\Delta h_b/h_b$
and $\delta h_b/h_b$ take the approximate numerical values $1/20$ and $-1/20$,
respectively, and $\kappa_b\approx\Delta h_b/h_b\approx 1/20$\footnote{For
comparison, in the $CP$-conserving Maximal Mixing
scenario $\kappa_b\simeq 1/200$, and in the Minimal Mixing scenario
$\kappa_b\simeq 1/400$. In both of these scenarios, $|\mu|= 200$~GeV.}. We denote the $CP$-even mass eigenstates by $h$ and $H$, where $M_h\leq M_H$. We can always identify $h$ with $H_1$, but due to strong radiative corrections $H$ can either be $H_2$ or $H_3$ depending on $M_{H^+}$ and $\tan\beta$.
The mass states are related to the gauge eigenstates by a mixing angle $\alpha$, which satisfies $(-\sin\alpha)=\mathcal{O}_{1h}$ and $\cos\alpha=\mathcal{O}_{1H}$. Furthermore, the pseudoscalar
effective couplings for these states vanish, and the scalar couplings
are rescaled relative to their tree level values. At tree level $g^S_{hb\bar{b}}$ and $g^S_{Hb\bar{b}}$ are given by $(-\sin\alpha/\cos\beta)$ and $(\cos\alpha/\cos\beta)$, respectively, and the rescaling factors are given by
\begin{eqnarray}
\frac{g^S_{hb\bar{b}}}{-\sin\alpha/\cos\beta}&=&
\frac{1-\kappa_b\cot\alpha}{1+\kappa_b\tan\beta}\nonumber\\
\frac{g^S_{Hb\bar{b}}}{\cos\alpha/\cos\beta}&=&
\frac{1+\kappa_b\tan\alpha}{1+\kappa_b\tan\beta}.
\label{reltree}
\end{eqnarray}
In Fig.~\ref{effhbb} we plot the squares
of $g^S_{hb\bar{b}}$ and $g^S_{Hb\bar{b}}$. 
In the large $M_{H^+}$ limit, $(-\sin\alpha)\rightarrow\cos\beta$
and $\cos\alpha\rightarrow\sin\beta$, so the $h$ scalar coupling converges
to the SM value. Similarly, in the small $M_{H^+}$ limit, the
$H_3$ scalar coupling becomes that of the SM.
Finally, in the non-standard Higgs limit (large $M_{H^+}$ for $H$, or
small $M_{H^+}$ for $h$) and for large $\tan\beta$, the effective coupling
of the non-standard Higgs approaches $\tan\beta/(1+\kappa_b\tan\beta)$, which
should be contrasted with the tree level behavior proportional to $\tan\beta$.
Although the coupling is still enhanced relative to the SM value, it is
suppressed relative to the tree level MSSM value.

It is important to be precise with terminology: as defined in the
Introduction, an SM-like Higgs is one with significant couplings to vector
bosons, not necessarily one with SM-like fermionic couplings. The
distinction is relevant, for example, at moderate values of $M_{H^+}$,
where the first two terms of the $h$ scalar coupling are comparable because of the factor of $\kappa_b$ in the
second term. For these values of $M_{H^+}$, $h$ can be simultaneously
SM-like and have an altered coupling to $b\bar{b}$. Of course, this is true
even for the tree level coupling, but the effective coupling is modified
relative to tree level by the factor in Eq.~\ref{reltree} which can generate
a significant suppression when $(-\cot\alpha) < \tan\beta$.

It is also important to observe that the $\tau^+\tau^-H_i$ effective coupling must take a similar form to that of
$b\bar{b}H_i$, but with radiative terms proportional to $\alpha_{1,2}$ instead of $\alpha_s$. The effects of the threshold corrections are therefore much smaller for the $\tau^+\tau^-$ coupling and can be qualitatively neglected.

When phases are introduced, $\mathcal{O}_{3i}$ can become nonzero for all
states. Then the third term of $g^S_{H_ib\bar{b}}$ in
Eq.~\ref{effcoup} is nonzero and proportional to the imaginary part
of $\kappa_b$, which is dominated by the imaginary part of $\Delta h_b/h_b$.
This term is $\tan\beta$-enhanced, so significant phases for $A_t$
and $M_{\tilde{g}}$ can affect the conclusions drawn above for vanishing
phases, potentially even countering the suppression effect in
$g^S_{H_ib\bar{b}}$.

\section{Tevatron Results}

Neutral Higgs states with SM-like couplings to gauge bosons are sought at
the Tevatron in associated production with a $W$ or $Z$ boson and
in gluon fusion channels, with the Higgs decay to $b\bar{b}$
providing the dominant decay channel in the former
case and Higgs decay to $W^+W^-$ providing the dominant channel in the
latter~\cite{Abazov:2008eb}--\cite{Aaltonen:2008ec}.
We compute projections for the expected upper limits on the signal from the combination of these channels at CDF and
D0 with 10 fb$^{-1}$ per channel and 0\%, 25\%, and 50\%
improvements in signal efficiencies. Neutral states with non-standard gauge
couplings are probed mainly in the inclusive $\phi\rightarrow\tau^+\tau^-$
channel~\cite{:2008hu,Abulencia:2005kq}, but also exclusively in associated production with bottom quarks,
with the Higgs decaying into either bottom quark or $\tau$ lepton pairs~\cite{Abazov:2008zz,CDF3b}.
For these channels we consider 7 and 10 fb$^{-1}$, but without
improvement in efficiency. The extension of the reach by efficiency
improvements can be easily estimated: since the signal scales approximately
with $\tan^2\beta$, any eventual improvement in efficiency will produce a further
extension of the reach in $\tan\beta$ by the square root of the efficiency
improvement.
Furthermore, the 95\% C.L. expected upper bound on the signal with
25\% improvements is essentially equivalent to the 90\% limit with no
improvements\footnote{This can be understood from the approximate formula for
the expected n$\sigma$ upper limit on the signal,
$s_{n\sigma}\approx n\sqrt{b}/(\epsilon\sqrt{L})$, where $L$ is the
luminosity, $\epsilon$ is the signal efficiency, and $b$ is the expected
background.}. The charged Higgs is sought in decays of the top quark~\cite{Collaboration:2009zh,Abulencia:2005jd}, $t\rightarrow H^+b$, with $H^+\rightarrow\tau^+\nu$ providing the dominant $H^+$ decay channel for $\tan\beta > 1$. We present the non-standard Higgs results in combination with those from the charged Higgs, since both particles can be classified together as strictly beyond-the-SM scalars. Finally, we combine the two classes of searches, SM-like and non-standard~+~charged, to derive
the strongest possible constraint on the MSSM Higgs parameter space. In all figures the shaded gray regions denote exclusion limits from
LEP~\cite{Schael:2006cr}, and solid black indicates theoretically
disallowed regions.

Note that the expected 95\% C.L. limits are obtained under the assumption that the data reflects only the average number of background events. If signal is also present and the data reflects the average value of signal+background for some point in the MSSM parameter space, the observed limit will be somewhat weaker, with an average value given by $R_{obs}\approx R_{exp}+1$, where $R$ is the upper bound on the signal normalized to the expected signal in the MSSM\footnote{Interestingly enough, the present combined Tevatron bounds on the SM Higgs given in Ref.~\cite{Collaboration:2009je} show an observed bound that differs by about 1 from the expected bound in the low mass range. At $M_h=115\mbox{ GeV}$, $R_{exp}=1.78$ and $R_{obs}=2.7$.}. For further discussion, see Appendix A.

\begin{figure}[!htbp]
\begin{center}
\begin{tabular}{cc}
\includegraphics[width=0.40\textwidth]{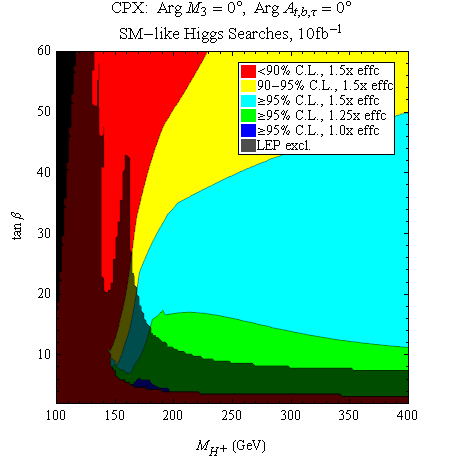} &
\includegraphics[width=0.40\textwidth]{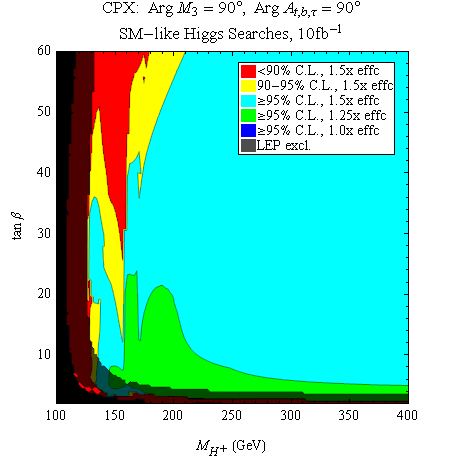} \\
\includegraphics[width=0.40\textwidth]{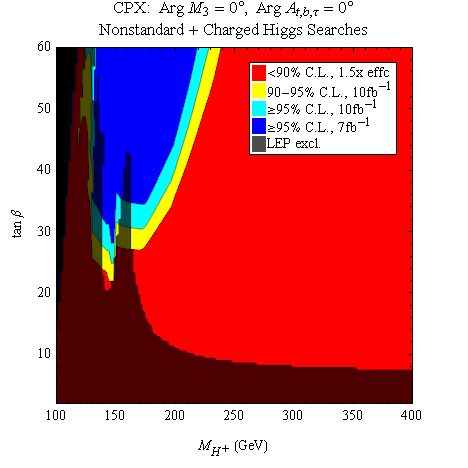} &
\includegraphics[width=0.40\textwidth]{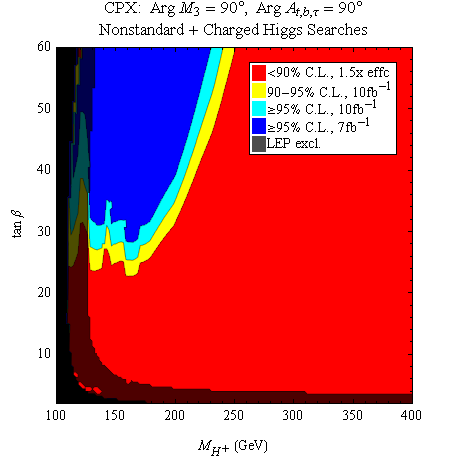} \\
\includegraphics[width=0.40\textwidth]{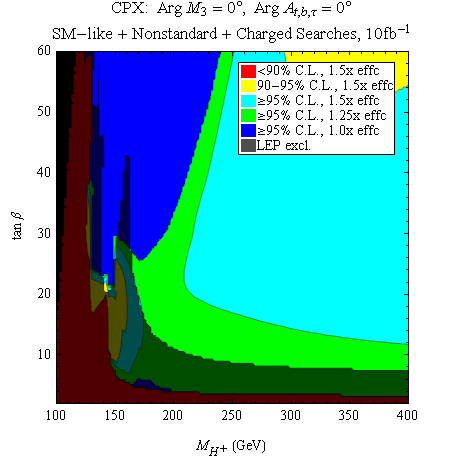} &
\includegraphics[width=0.40\textwidth]{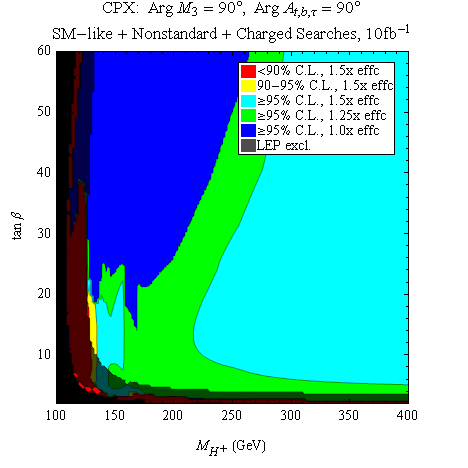}
\end{tabular}
\caption{Projected Tevatron exclusion contours at 90\% and 95\% C.L. in the
CPX scenario with $CP$-violating phases $\arg(A_{t,b,\tau})=0^{\circ}$,
$\arg(M_{\tilde{g}})=0^{\circ}$ (\textit{left}) and
$\arg(A_{t,b,\tau})=90^{\circ}$, $\arg(M_{\tilde{g}})=90^{\circ}$
(\textit{right}). Row 1 gives results for SM-like Higgs searches,
row 2 includes only the non-standard Higgs searches, and row 3
gives the combined constraints.}
\label{CPX0000}
\end{center}
\end{figure}

In this work we will consider three sets of phases for
$A_{t,b,\tau}$ and $M_{\tilde{g}}$: $(0^{\circ},0^{\circ})$,
$(90^{\circ},90^{\circ})$, and $(140^{\circ},140^{\circ})$. In the course
of our study we examined other values of the $CP$-violating phases, including
the departure from setting common phases for $A_{t,b,\tau}$ and
$M_{\tilde{g}}$. However, the results were qualitatively similar, and
in particular, all unique features of interest also appeared in one or more
cases discussed here.


We consider first the case without explicit $CP$-violation, in order
to understand features which are independent of the phases.  The results
from SM-like searches, non-standard + charged Higgs search channels, and the combination are
given in the first column of Fig.~\ref{CPX0000}. The decoupling limit
is probed at 95\% C.L. with a 50\% improvement in signal efficiency for the
SM-like search channels. In this limit $M_{H_1}\approx 121\mbox{ GeV}$ with
the top mass set to $173.1\mbox{ GeV}$. Smaller efficiency improvements more
readily probe the region of lower $\tan\beta$,
where $M_{H_1}\lesssim 121\mbox{ GeV}$ and the SM Higgs constraint is stronger.

Here a word about experimental mass resolution is in order. If the mass
difference between two Higgs states is under a certain finite threshold,
the detectors cannot resolve the particles. Consequently, their statistical
significances should be added directly rather than in quadrature,
leading to a stronger limit.  We take a representative experimental mass
resolution of $10\mbox{ GeV}$. The main effect of this approximate treatment appears as a discontinuous
spike in the range $M_{H^+}\approx 135-155\mbox{ GeV}$ of the non-standard
Higgs search constraint. The reason for this behavior is that in this range of $M_{H^+}$ and for moderate
to large $\tan\beta$, radiative corrections drive a $CP$-even Higgs mass to within $10\mbox{ GeV}$ of the $CP$-odd Higgs mass. For slightly lower or higher values of $M_{H^+}$ the approximate degeneracy is lifted. In all subsequent figures of this work, we include the finite mass resolution effects.

\begin{figure}[!htbp]
\begin{center}
\begin{tabular}{cc}
\includegraphics[width=0.40\textwidth]
{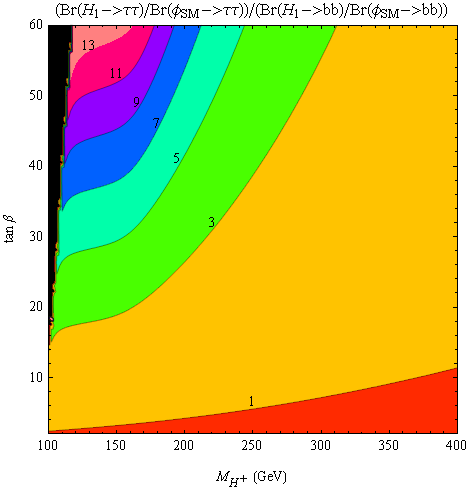} &
\includegraphics[width=0.40\textwidth]
{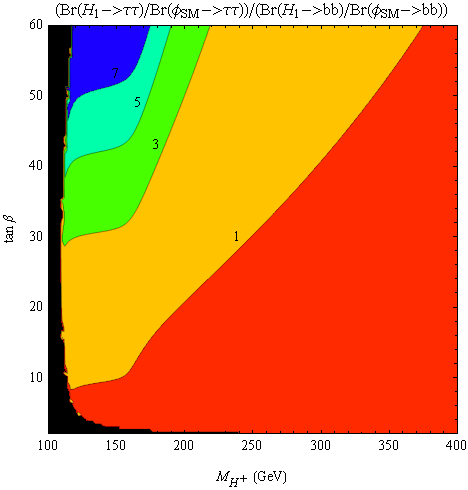} \\
\end{tabular}
\includegraphics[width=0.40\textwidth]
{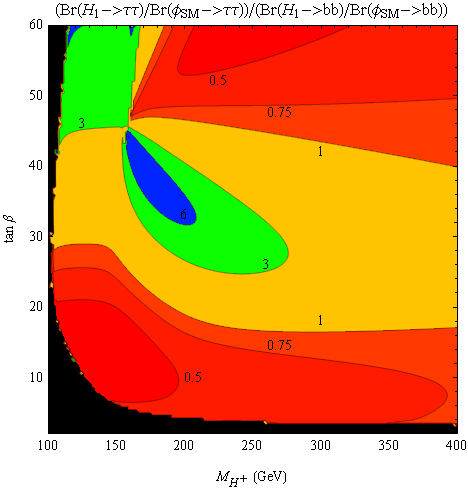} \\
\caption{The ratio of $H_1$ branching ratios into $\tau^+\tau^-$
and $b\bar{b}$, normalized to their SM values, for vanishing
phases (\textit{upper left}), for $\arg(A_{t,b,\tau})=90^{\circ}$,
$\arg(M_{\tilde{g}})=90^{\circ}$ (\textit{upper right}), and for $\arg(A_{t,b,\tau})=140^{\circ}$,
$\arg(M_{\tilde{g}})=140^{\circ}$ (\textit{bottom}).}
\label{h1tth1bb}
\end{center}
\end{figure}

Contrary to the $CP$-conserving cases analyzed in Ref.~\cite{Draper:2009fh},
the SM-like Higgs search constraints become weaker with moderate,
decreasing $M_{H^+}$. At tree level, the increase in $\mathcal{O}_{11}$
significantly enhances the coupling of the SM-like Higgs to down-type fermions
for smaller $M_{H^+}$, typically leading to a large region just above the
intense coupling regime where the constraint is stronger than in the
decoupling limit (see, for example, the Maximal Mixing scenario examined in
Ref.~\cite{Draper:2009fh}). This feature does not appear in the CPX scenario
in the absence of phases. The reason is that, as discussed before and
shown in Fig.~\ref{effhbb},
the $\kappa_b$ threshold corrections in Eq.~\ref{effcoup} are significant due
to the large value of $\mu$ taken in CPX. These corrections suppress the
$b\bar{b}H_1$ effective coupling relative to the tree level value, and therefore $Br(H_1\rightarrow b\bar{b})$ is
decreased while $Br(H_1\rightarrow \tau^+\tau^-)$ is increased in this region.
This can be seen from the upper left plot of Fig.~\ref{h1tth1bb}, where we
show the ratio of the $\tau^+\tau^-$ to $b\bar{b}$ branching ratios, each
normalized to their SM values.

The constraint from the non-standard Higgs search in the $\tau^+\tau^-$
inclusive channel is similar to what is obtained in other benchmark scenarios
geared towards the $CP$-conserving MSSM. At tree level the non-standard Higgs
has a $\tan\beta$ enhanced coupling to $\tau^+\tau^-$ that is not subject to
large radiative corrections, and it is light enough to be produced for low to
moderate $M_{H^+}$. As a result the constraint is significant in this region.
The radiative increase in the $\tau^+\tau^-$ branching fraction due
to the suppression of $b\bar{b}$ is mostly compensated by the threshold
suppression of the non-standard Higgs coupling to bottom quarks, which enters
in both of the dominant production mechanisms of $b\bar{b}$ fusion and
gluon fusion through a bottom loop~\cite{Carena:2005ek}. $H_1$ becomes highly non-standard for moderate to large $\tan\beta$ and $M_{H^+}\lesssim 150\mbox{ GeV}$.
Therefore the limit comes mostly from $H_1$ and the $CP$-odd Higgs for $M_{H^+}\lesssim 150\mbox{ GeV}$, and
from the $CP$-odd and heavy $CP$-even Higgs for larger $M_{H^+}$. The non-standard constraint
has the virtue of mostly filling the dip in the LEP constraint
at $M_{H^+}\approx 140\mbox{ GeV}$, which was due to a marginal excess
in the LEP data around $M_{H_1}\approx 90\mbox{ GeV}$. The charged Higgs searches from top decays become also
relevant in this region of parameters, although in the absence of phases they do not independently reach 95\% C.L. due to the large bottom coupling suppression in this scenario~\cite{Carena:1999py}.

We stress the fact that neither the SM-like nor the non-standard~+~charged Higgs
search is sufficient to reach the entire plane even with significant improvement.
However, each search is most effective in the region where the other is
weakest, providing excellent complementarity and strongly motivating a
statistical combination. In the bottom left plot of Fig.~\ref{CPX0000},
we demonstrate that the combination of SM-like searches with 50\%
improvements in signal efficiency and non-standard~+~charged search channels is
sufficient to cover the entire region previously unprobed by LEP.

Now we consider the effects of $CP$-violation, setting
$\arg(A_{t,b,\tau})=\arg(M_{\tilde{g}})=90^{\circ}$. The results are
presented in the right-hand column of Fig.~\ref{CPX0000}. In the
decoupling limit, $H_1$ is still SM-like; however,
for $M_{H^+}\lesssim 160\mbox{ GeV}$, $H_1$ becomes mostly $CP$-odd.
$H_2$ is $CP$-odd in the decoupling limit, but transitions rapidly to become
SM-like around $M_{H^+}\approx 150\mbox{ GeV}$, and
finally acquires non-standard couplings to gauge bosons for
$M_{H^+}\lesssim 135\mbox{ GeV}$.
The fast transitions create a region of increased sensitivity compared to the $CP$-conserving case centered around
$M_{H^+}\approx 150\mbox{ GeV}$ and stretching from low to moderate
$\tan\beta$. Although $g_{ZZH_2}$ increases with $\tan\beta$, the $H_2\rightarrow b\bar{b}$ branching ratio suppression limits the height of the region. The region ends sharply at $M_{H^+}\approx 130\mbox{ GeV}$, where $H_3$ has become SM-like but the opening of the $H_3\rightarrow H_1H_1$ channel heavily reduces the $H_3\rightarrow b\bar{b}$ branching ratio. As before the LEP constraint from $e^+e^-\rightarrow H_1H_2\rightarrow4b,2b2\tau$ takes over for lower values of $M_{H^+}$.

The phase for $M_{\tilde{g}}$ influences the Higgs masses more mildly than phases
for the trilinear couplings because it enters the mass matrix only at
the 2-loop level. Nonetheless, it strengthens the SM-like constraint around
$M_{H^+}\approx 200\mbox{ GeV}$, primarily by counteracting the threshold
suppression of $g^S_{H_ib\bar{b}}$ as discussed in Section 2 and leading to
a $g^S_{H_ib\bar{b}}$ that is enhanced over the tree level value in this
region. Correspondingly the feature familiar from the Maximal Mixing scenario,
which we noted earlier was absent in CPX with vanishing phases, has begun to
reemerge in the small $M_{H^+}$, moderate $\tan\beta$ region. For comparison with the case of vanishing phases,
in the upper right plot of Fig.~\ref{h1tth1bb} we again present the ratio of
the $b\bar{b}$ to $\tau^+\tau^-$ branching ratios, each normalized to their
SM values, now in the presence of the $90^{\circ}$ phases.

The non-standard~+~charged Higgs reach is similar to the case without phases. The two
spikes, most visible on the combined plot, are the result of the mass
resolution prescription discussed above. Below
$M_{H^+}=150\mbox{ GeV}$, $H_3$ and $H_2$ are within $10\mbox{ GeV}$;
above, $H_2$ and $H_1$ share this property. From the final figure, it is
again evident that the combination of channels is essential to cover nearly the entire plane at 95\% C.L.

We note that with $90^{\circ}$ phases there appears a small hole in the
LEP coverage at low $M_{H^+}$ and low $\tan\beta$ adjacent to the
theoretically disallowed region, which is unprobed by any of the
Tevatron search channels.
The presence of this hole was
first discussed in Ref.~\cite{Carena:2002bb} and is generated by the possible
decay of the SM-like Higgs boson into a pair of $H_1$'s, which acquire a
significant $CP$-odd Higgs component. The hole is discussed in detail in Ref.~\cite{Williams:2007dc}
and channels which may help to cover it at hadron colliders are studied in
Ref.~\cite{Kaplan:2009qt}--\cite{Chang:2005ht}.
Our results are qualitatively consistent with the LEP experiment plots
from Ref.~\cite{Schael:2006cr}\footnote{In our plots the hole is somewhat
smaller than in Ref.~\cite{Schael:2006cr} due to the finite grid size in our scan and
the approximations we used to implement the LEP constraints.}.
In the $M_{H^+}$ coordinates this hole appears as a very small region;
however, it covers a significant portion of the range
$M_{H_1}\lesssim 45\mbox{ GeV}$, and indicates that a light Higgs scenario
phenomenologically similar to what has been proposed in the context of the
NMSSM~\cite{Dermisek:2005gg} has still not been fully ruled out in the
MSSM. In this hole the decays $H_2\rightarrow H_1H_1\rightarrow 4b,4\tau$
considered by LEP are significant; however, since $\tan\beta$ is small,
$H_2$ is of mixed composition and its dominant production mechanisms are
suppressed. Recently, the authors of Ref.~\cite{Cranmer} reanalyzed the ALEPH data, extending the reach of the $4\tau$ channel to higher values of $M_{H_2}$. However, this channel still only covers a small subset of this hole, because it is only efficient for $2m_{\tau}<M_{H_1}<2m_b$, and the $ZZH_2$ coupling is typically suppressed in this region.

\begin{figure}[!htbp]
\begin{center}
\begin{tabular}{cc}
\includegraphics[width=0.40\textwidth]
{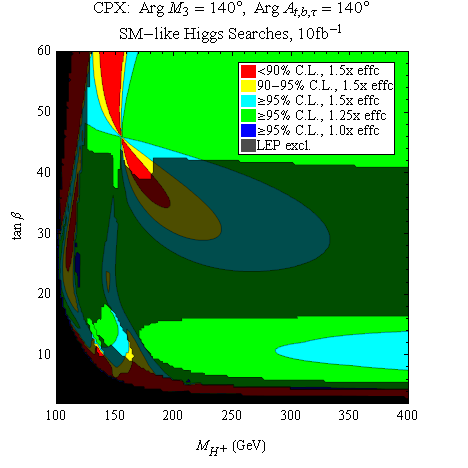} \\
\includegraphics[width=0.40\textwidth]
{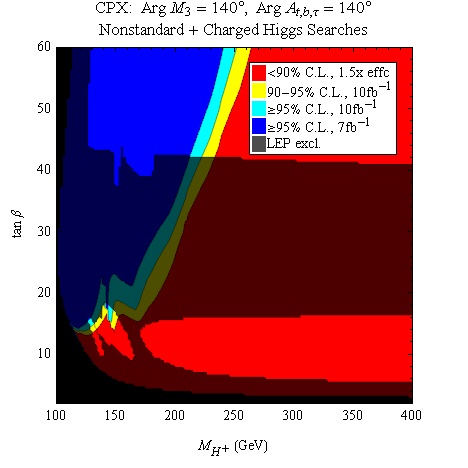} \\
\includegraphics[width=0.40\textwidth]
{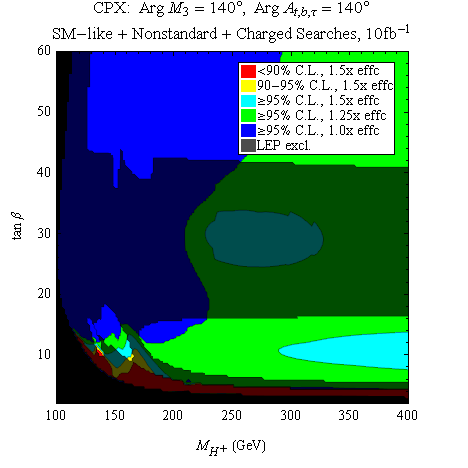} \\
\end{tabular}
\caption{Projected Tevatron exclusion contours at 90\% and 95\% C.L.
in the CPX scenario with $CP$-violating phases
$\arg(A_{t,b,\tau})=140^{\circ}$, $\arg(M_{\tilde{g}})=140^{\circ}$.
Row 1 gives results for SM-like Higgs searches, row 2 includes only the
non-standard Higgs searches, and row 3 gives the combined constraints.}
\label{CPX140140}
\end{center}
\end{figure}

Finally, we consider a case with larger values for the phases, with
projections given in Fig.~\ref{CPX140140}. For
$\arg(A_{t,b,\tau})=140^{\circ}$ and $\arg(M_{\tilde{g}})=140^{\circ}$,
the coverage appears more unusual than in previous cases.
Near $M_{H^+}\approx 160\mbox{ GeV}$ and $\tan\beta\approx 45$, $M_{H_1}$
and $M_{H_2}$ become degenerate. Around this point there is sufficient variation in the mixing
matrix so that in one direction the SM Higgs becomes strongly up-type, suppressing the $\tau^+\tau^-$ width, while in another direction the different terms in Eq.~\ref{effcoup}
interfere destructively and suppress the $b\bar{b}$ width. The result is demonstrated in the bottom plot of Fig.~\ref{h1tth1bb}: there exist both large areas of parameter space
where the branching ratio to $b\bar{b}$ is suppressed while $\tau^+\tau^-$ is
enhanced, and large areas where the reverse occurs. The former leads to the
unprobed red stripes in the SM-like searches, centered on the degeneracy point
at moderate to large $\tan\beta$, but the latter causes the coverage to be
extended to a 95\% C.L. limit with only a 25\% efficiency improvement for
large $\tan\beta$ and moderate to large $M_{H^+}$.

The LEP coverage traces
out a region of moderate $\tan\beta$ where sbottom corrections to
$M_{H_1}$ are maximized~\cite{Carena:2002bb}. These corrections are negative and proportional
to $|h_b|^4$, where $h_b$ is the effective bottom quark Yukawa coupling.
The modulus depends on the $CP$-violating phases in such a way that
if $\cos\phi<0$ (assuming a common phase $\phi$ for $A_t$
and $M_{\tilde{g}}$), then $M_{H_1}$ has a local minimum
at $\cot\beta=-|\Delta h_b/h_b|\cos\phi$. As we noted earlier,
$|\Delta h_b/h_b|\approx -1/20$, so around
$\tan\beta\approx 25$ the LEP constraints are stronger than
for either lower or higher values of $\tan\beta$.

In the $(140^{\circ},140^{\circ})$ case the utility of the non-standard
and SM-like combination is particularly manifest. Most of parameter space
can be covered by SM-like searches, with the exception of the stripes where
the $b\bar{b}$ branching ratio of the SM-like Higgs is suppressed. The
stripes occur precisely in the low $M_{H^+}$, large $\tan\beta$ region where
the non-standard search is most effective. This effect is not unique to the
MSSM with $CP$-violation; a similar suppression of the $b\bar{b}$ channels
in this region and the corresponding complementarity of coverage exists in
the small-$\alpha_{eff}$ scenario studied in Refs.~\cite{Draper:2009fh},\cite{Carena:2002qg}--\cite{Hahn:2006my}.

\section{LHC Results}

For the LHC we examine the discovery reach for an integrated luminosity of 3 fb$^{-1}$ in all channels. For the case of $140^{\circ}$ phases, which is the most difficult to probe, we also present the reach for 30 fb$^{-1}$. Table~\ref{Tch} lists the search channels we use from CMS, taken from Refs.~\cite{Ball:2007zza,CMSWW1,CMSWW2}, and the channels from ATLAS, given in Ref.~\cite{Aad:2009wy}. The searches for $\phi\rightarrow\tau^+\tau^-,\mu^+\mu^-$ in association with bottom quarks are probes of the non-standard Higgs; the rest are SM-like Higgs channels. The experimental studies of the non-standard channels present the expected $5\sigma$ contour in the $(M_{A},\tan\beta)$ plane in $CP$-conserving benchmark scenarios. However, to a good approximation the production cross section of the non-standard Higgs via bottom quark fusion is independent of the benchmark values, and scales as $\tan^2\beta$ in most of parameter space. Therefore we employ this scaling relation to extend the experimental analyses to expected significances on the full $(M_{H^+},\tan\beta)$ plane in CPX. Furthermore, for simplicity we omit channels which probe the charged Higgs, where the expected discovery region is mostly\footnote{Small regions at low $\tan\beta\lesssim 5$ that are not reached by the non-standard Higgs searches may be probed by the charged Higgs channels. However, these regions tend to be either theoretically disallowed or already excluded by LEP.} a subset of the region probed by the non-standard neutral Higgs at high significance.

\begin{table}[!htbp]
\begin{center}
\begin{tabular}{|c|c|c|}
  \hline
  Experiment & Production & Decay \\
  \hline
  \hline
  & Weak Boson Fusion & $\phi\rightarrow\tau^+\tau^-$ \\
  \cline{2-3}
  & Inclusive & $\phi\rightarrow\gamma\gamma,W^+W^-,ZZ$ + tagged\\
  & & leptons from each $W$ and $Z$\\
  \cline{2-3}
  CMS & Weak Boson Fusion & $\phi\rightarrow W^+W^-$ with one tagged\\
  & & lepton and two jets \\
  \cline{2-3}
  & $t\bar{t}\phi$ & $\phi\rightarrow b\bar{b}$ \\
  \cline{2-3}
  & $b\bar{b}\phi$ & $\phi\rightarrow\tau^+\tau^-,\mu^+\mu^-$ \\
  \hline
  & Weak Boson Fusion + Gluon Fusion & $\phi\rightarrow W^+W^-$ \\
  \cline{2-3}
  & Inclusive & $\phi\rightarrow\gamma\gamma,ZZ$ \\
  \cline{2-3}
  ATLAS & Weak Boson Fusion & $\phi\rightarrow\tau^+\tau^-$ \\
  \cline{2-3}
  & $t\bar{t}\phi$ & $\phi\rightarrow b\bar{b},W^+W^-$ \\
  \cline{2-3}
  & W$\phi$ & $\phi\rightarrow b\bar{b}$ \\
  \cline{2-3}
  & $b\bar{b}\phi$ & $\phi\rightarrow\tau^+\tau^-,\mu^+\mu^-$ \\
  \hline
\end{tabular}
  \caption{Search channels from the LHC employed in this study.}
    \label{Tch}
\end{center}
\end{table}

\begin{figure}[!htbp]
\begin{center}
\begin{tabular}{cc}
\includegraphics[width=0.40\textwidth]{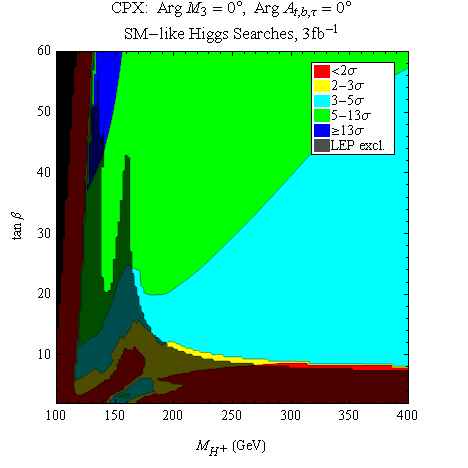} &
\includegraphics[width=0.40\textwidth]{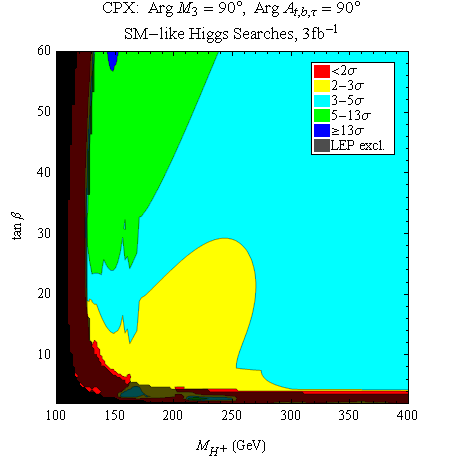} \\
\includegraphics[width=0.40\textwidth]{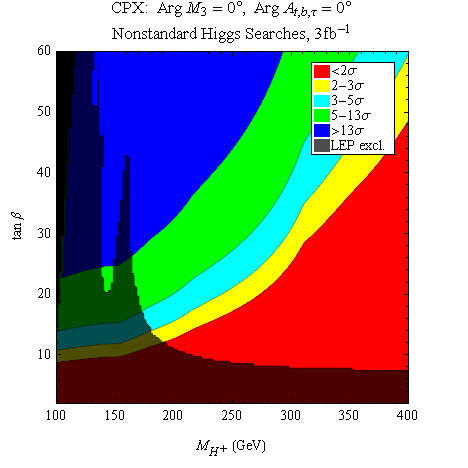} &
\includegraphics[width=0.40\textwidth]{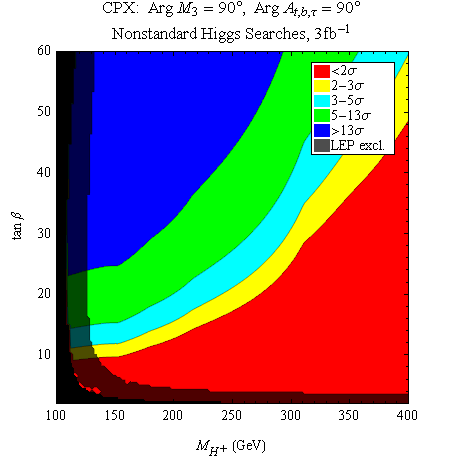} \\
\includegraphics[width=0.40\textwidth]{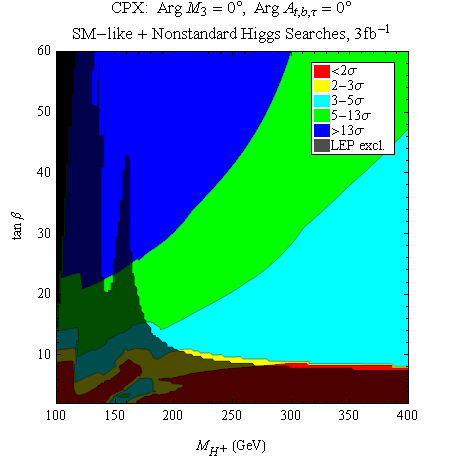} &
\includegraphics[width=0.40\textwidth]{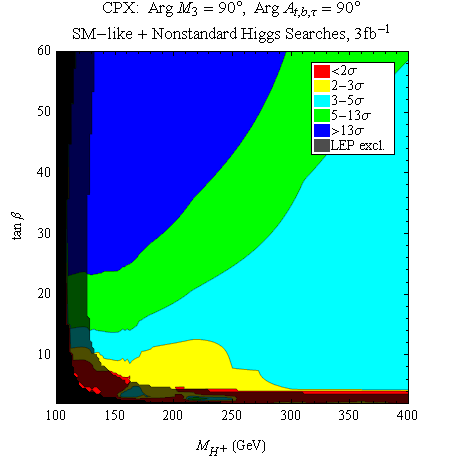} \\
\end{tabular}
\caption{Projected LHC significance contours for 3 fb$^{-1}$ in the CPX scenario with phases $\arg(A_{t,b,\tau})=0^{\circ}$, $\arg(M_{\tilde{g}})=0^{\circ}$ (\textit{left}), and $\arg(A_{t,b,\tau})=90^{\circ}$, $\arg(M_{\tilde{g}})=90^{\circ}$ (\textit{right}). Row 1 gives results for SM-like Higgs searches, row 2 includes only the non-standard Higgs searches, and row 3 gives the combined constraints.}
\label{LHC0000}
\end{center}
\end{figure}

\begin{figure}[!htbp]
\begin{center}
\begin{tabular}{cc}
\includegraphics[width=0.40\textwidth]{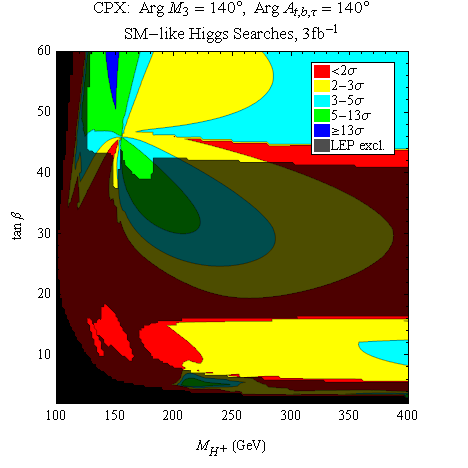} &
\includegraphics[width=0.40\textwidth]{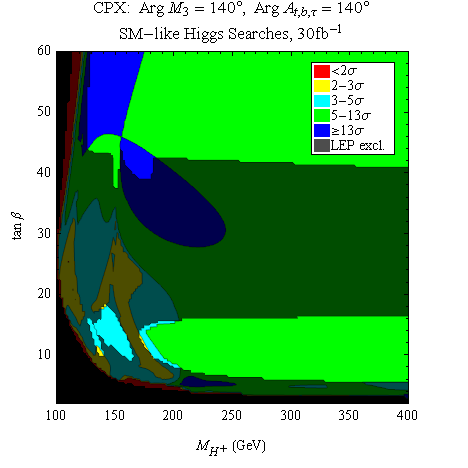} \\
\includegraphics[width=0.40\textwidth]{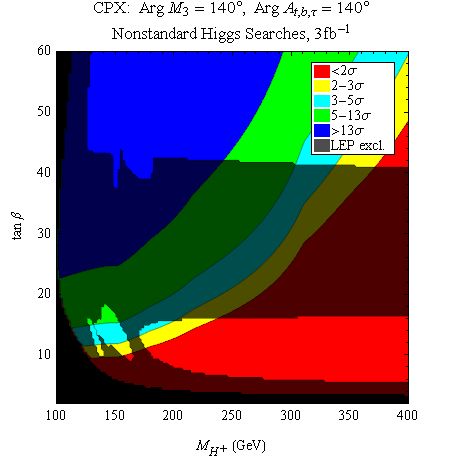} &
\includegraphics[width=0.40\textwidth]{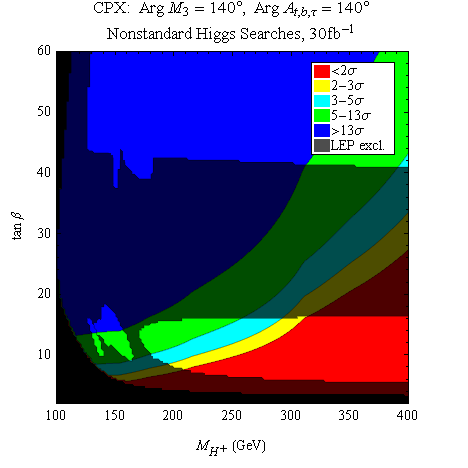} \\
\includegraphics[width=0.40\textwidth]{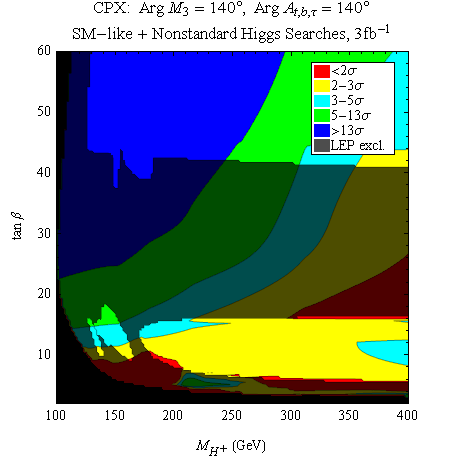} &
\includegraphics[width=0.40\textwidth]{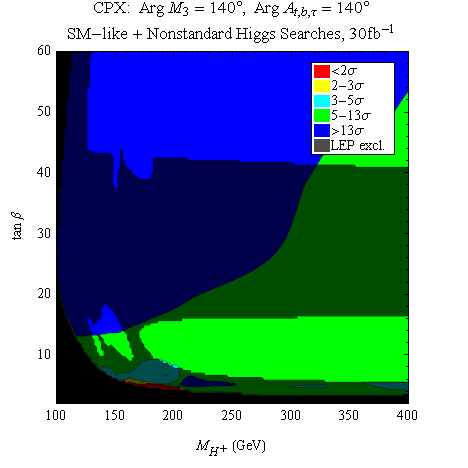} \\
\end{tabular}
\caption{Projected LHC significance contours in the CPX scenario with phases $\arg(A_{t,b,\tau})=140^{\circ}$, $\arg(M_{\tilde{g}})=140^{\circ}$, for 3~fb$^{-1}$ (\textit{left}) and 30~fb$^{-1}$ (\textit{right}). Row 1 gives results for SM-like Higgs searches, row 2 includes only the non-standard Higgs searches, and row 3 gives the combined constraints.}
\label{LHC140140}
\end{center}
\end{figure}

The left-hand column of Fig.~\ref{LHC0000} gives the projections
for $(\arg(A_{t,b,\tau}),\arg(M_{\tilde{g}}))=(0^{\circ},0^{\circ})$, and
the right-hand column contains the case with phases set to $(90^{\circ},90^{\circ})$.
Fig.~\ref{LHC140140} examines the reach with $(140^{\circ},140^{\circ})$, where
the left-hand column assumes 3~fb$^{-1}$ and the right-hand column uses 30~fb$^{-1}$.

The dominant SM-like search channels in all cases are
$\phi\rightarrow\tau^+\tau^-$ with weak boson fusion (WBF)
production~\cite{Rainwater:1999sd,Plehn:1999nw,Plehn:1999xi} and the inclusive search for $\phi\rightarrow\gamma\gamma$~\cite{Carena:1999bh}.
Since the $\tau^+\tau^-$ channel relies on the coupling of the Higgs state
to vector bosons, it is indeed an SM-like search; however, the strong
enhancement of the $\phi\rightarrow\tau^+\tau^-$ branching ratio evident in Fig.~\ref{h1tth1bb}
implies that this search is strongest in the intense
coupling regime, where multiple states may have simultaneously moderate
gauge couplings and enhanced branchings to $\tau^+\tau^-$.
However, as shown in Fig.~\ref{h1tth1bb}, the introduction of $90^{\circ}$
phases reduces the
$\tan\beta$-enhanced suppression of the coupling to bottom quarks,
preventing the $\tau^+\tau^-$ branching fraction from becoming as large as
in the case with vanishing phases. This makes the $\tau^+\tau^-$ channel
relatively weaker than in the case without $CP$-violation, and in particular
causes the low $\tan\beta$, large $M_{H^+}$ region to be unprobed by this
channel. Instead, the region is covered by the $\phi\rightarrow\gamma\gamma$
search, which can probe the decoupling limit
at $2-3\sigma$ with 3 fb$^{-1}$.

\begin{figure}[!htbp]
\begin{center}
\begin{tabular}{cc}
\includegraphics[width=0.40\textwidth]{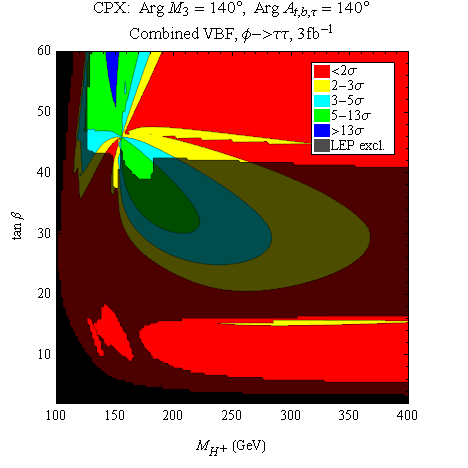} &
\includegraphics[width=0.40\textwidth]{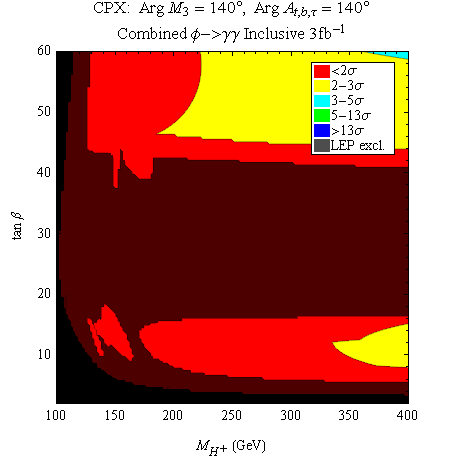} \\
\end{tabular}
\caption{Projected LHC significance contours in the CPX scenario with phases $\arg(A_{t,b,\tau})=140^{\circ}$, $\arg(M_{\tilde{g}})=140^{\circ}$ and 3~fb$^{-1}$, derived from the WBF $\phi\rightarrow\tau^+\tau^-$ channel  (\textit{left}) and the inclusive $\phi\rightarrow\gamma\gamma$ channel (\textit{right}).}
\label{LHC140indiv}
\end{center}
\end{figure}

As discussed previously for the Tevatron, in the $(140^{\circ},140^{\circ})$
scenario there appears a region $M_{H^+}\gtrsim 200\mbox{ GeV}$
and $\tan\beta\gtrsim 45$ where the $H_1\rightarrow\tau^+\tau^-$ branching ratio is significantly suppressed. The individual limits from the $\tau^+\tau^-$ and $\gamma\gamma$ channels are given for the case of 3~fb$^{-1}$ in Fig.~\ref{LHC140indiv}. The $H_1\rightarrow \gamma\gamma$ channel
is strong enough to probe this region at the $2-3\sigma$ level
with 3 fb$^{-1}$, or the $6-8\sigma$ level with 30 fb$^{-1}$.
However, there are also significant holes in the LEP coverage for lower values of $\tan\beta$ across most of the range of $M_{H^+}$, and in these regions, the branching fractions of the SM-like Higgs to $\tau^+\tau^-$ and $\gamma\gamma$ are always suppressed. Even after combining the channels there remain regions centered at $M_{H^+}\approx 150\mbox{ GeV}$ and $M_{H^+}\approx 200\mbox{ GeV}$ that may be probed at less than $2\sigma$ with 3 fb$^{-1}$. Furthermore, when we increase the luminosity to 30 fb$^{-1}$, $5\sigma$ discovery reach contour still does not cover the region around $150\mbox{ GeV}$ and $\tan\beta\approx 10-15$.

In all sets of phases, the combination of SM-like and non-standard channels considerably strengthens the statistical significance of the LHC results with 3 fb$^{-1}$. Note that in the context of this combination, a $5\sigma$ significance does not necessarily imply a resonance peak in the data with this significance corresponding to any Higgs state; for example, it could be generated by only a $3\sigma$ excess coming from the SM-like Higgs combined with a $4\sigma$ excess from the nonstandard Higgs. The correct interpretation is that taken together, these excesses indicate an exclusion of the background-only hypothesis at $5\sigma$. Of course, it is desirable to eventually discover the particles individually and measure their properties. This will take longer, but can be achieved with a higher integrated luminosity: we have checked that when the data set approaches 30 fb$^{-1}$, it is no longer necessary to go beyond the SM-like searches in order to probe the whole plane at 5$\sigma$ in
the cases of vanishing and $90^{\circ}$ phases. However, as mentioned previously, for $140^{\circ}$ phases there is still a small region at low $\tan\beta$ and low $M_{H^+}$ which is unconstrained by LEP and is not probed at 5$\sigma$ by the SM-like
channels with 30 fb$^{-1}$. This demonstrates that even with a considerable amount of data the combined analysis of standard and non-standard Higgs searches may still remain relevant.

\begin{figure}[!htbp]
\begin{center}
\begin{tabular}{cc}
\includegraphics[width=0.40\textwidth]{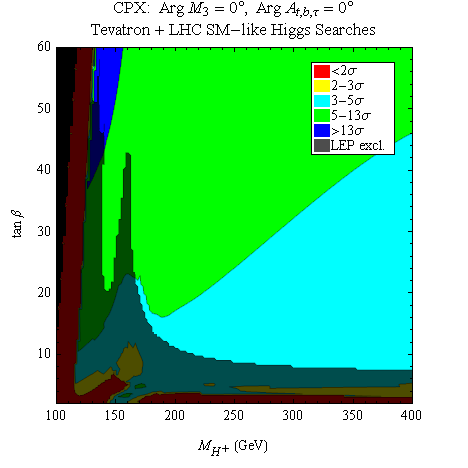} &
\includegraphics[width=0.40\textwidth]{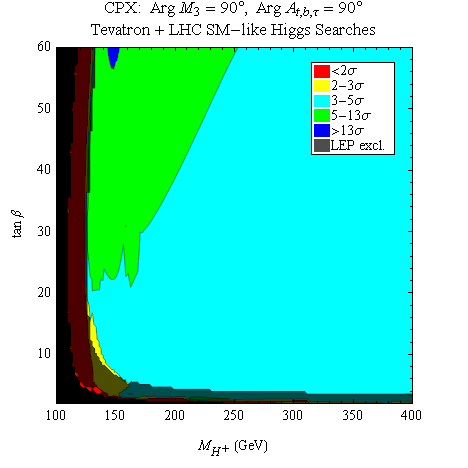} \\
\end{tabular}
\includegraphics[width=0.40\textwidth]{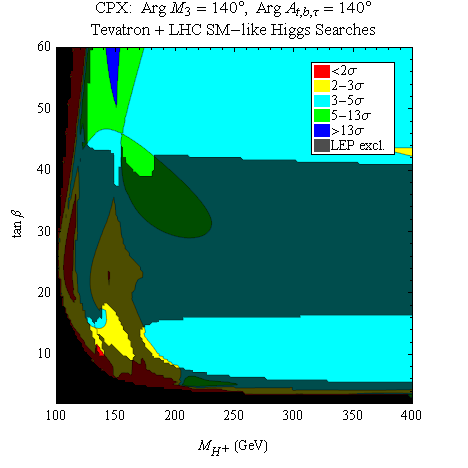}
\caption{Projected discovery significance contours from the combination of 10 fb$^{-1}$ + 50\% efficiency-improved Tevatron data and 3 fb$^{-1}$ LHC data, using SM-like search channels in the CPX scenario for each of the previously considered sets of phases.}
\label{clashofthetitans}
\end{center}
\end{figure}

It is interesting to note that whereas the strongest LHC SM-like Higgs searches are in the $\tau^+\tau^-$ and $\gamma\gamma$ channels, the Tevatron searches are mainly sensitive to the $b\bar{b}$ branching ratio. This generates complementarity in the coverage offered by the two colliders.
In Fig.~\ref{clashofthetitans} we give the discovery significance plots for each set of phases obtained by combining the projected reach of the SM-like search channels with 3~fb$^{-1}$ of data from the LHC and 10~fb$^{-1}$ + 50\% efficiency
improvements from the Tevatron\footnote{In our analysis of the Tevatron in Section 3, we test the signal~+~background hypothesis for the
purpose of limit-setting, whereas for the LHC we calculate discovery significance, which is a test of the background-only hypothesis.
However, in the limit of large backgrounds, the discovery significance at the Tevatron is inversely proportional to the
the upper bound derived on the signal. Further discussion of this approximation and the combination of channels is given in Appendix A.}. Particularly in the cases with nonzero phases, there are large regions where the LHC coverage is at the $2-3\sigma$ level or lower. In combination with the Tevatron the potential reach achieves $3\sigma$ on most of the plane. As discussed in the Introduction, depending on the data-taking rate of the LHC when it reaches a few fb$^{-1}$, this complementarity suggests that it may be worthwhile to perform the combination. Further examples of Tevatron-LHC combinations are presented in Appendix B for standard $CP$-conserving benchmark scenarios.

\section{Conclusions}

In this work we have analyzed in detail the Tevatron and LHC reach for the
Higgs sector in the MSSM with explicit $CP$-violating phases,
providing the most up-to-date projections for the exclusion and
discovery potential of these machines in this scenario.
Our primary goals
regarding the Tevatron search analyses
were to outline improvement factors necessary to probe
most of the parameter space, and to exhibit the considerable complementarity
offered by search channels applicable for Higgs states with SM-like gauge
couplings, and channels relevant for charged Higgs bosons and
neutral Higgs states with negligible gauge couplings
but enhanced couplings to fermions.
We find that taken in statistical
combination these two classes of Higgs searches can probe the
entire parameter plane in a benchmark scenario that exhibits strong
$CP$-violating  influence on the MSSM Higgs sector.

The low luminosity LHC offers complementary capabilities to the Tevatron.
The SM-like Higgs searches tend to be more efficient in the regions
where the Higgs decay branching ratio into bottom quark pairs is suppressed,
a region that becomes quite difficult for searches at the Tevatron. With this in mind,
we have examined the possibility of combining SM-like Higgs search
results from the Tevatron and LHC, and found that it can serve to
probe nearly the whole parameter space at more than 2$\sigma$ in regions in
which neither the Tevatron nor a low-luminosity LHC can do so by
standard Higgs channels alone. Therefore,
in these cases the combination may be a worthwhile exercise before the LHC has run for the
few years necessary to collect a large data sample. Furthermore, as
happens with the Tevatron, the combination of standard and
non-standard Higgs searches at the LHC allows essentially full coverage of the parameter space
at more than 2$\sigma$ with low integrated luminosity.  Finally, we showed
that this combination remains a powerful tool to disclose the full LHC discovery potential at
higher luminosities.

In our treatment of the CP-violating MSSM in this study, and previously in the CP-conserving case examined in Ref.~\cite{Draper:2009fh}, we took specific representative values of the soft supersymmetry breaking parameters and the Higgsino mass parameter at the electroweak scale in order to study the reach of collider experiments for certain generic or interesting features of the MSSM Higgs sector. Constraints on the Higgs sector that can be derived from flavor physics or dark matter considerations are very sensitive to the supersymmetry breaking scale and to precise relations between the soft supersymmetry breaking parameters of the different generations as well as the gaugino masses. For this reason,
we restricted our analysis to the direct collider constraints on the Higgs sector alone. However, considering the renormalization group evolution of these parameters, our procedure can be applied to study particular UV-embeddings of the MSSM, including specific models of the mediation of supersymmetry breaking into the MSSM. The potential collider reach for these models is interesting in its own right, and furthermore requires the application of flavor and dark matter constraints. We leave the investigation of these interesting possibilities to future work.


\section*{Acknowledgments}

We would like to thank S. Farkas, T. Junk, B. Kilminster, and M. Oreglia for helpful discussions on statistical methods. Work at ANL is supported in part by the U.S. Department of Energy (DOE), Div.~of HEP, Contract DE-AC02-06CH11357. Work at EFI is supported in part by the DOE through Grant No. DE-FG02- 90ER40560.  T.L. is also supported by the
Fermi-McCormick Fellowship. This work was supported in part by the DOE under Task TeV of contract DE-FGO2-96-ER40956.

\newpage

\appendix

\renewcommand{\thesection}{Appendix \Alph{section}}

\setcounter{equation}{0}

\section{Statistical Methods and Approximations}

\renewcommand{\theequation}{\Alph{section}.\arabic{equation}}

In this Appendix we review the statistics and approximations underpinning the calculations of exclusion limits and discovery used in this work. We begin with the familiar, intuitive understanding of these quantities, valid in the Gaussian limit. We then sketch how they are derived from the precise formulae used by the LHC collaborations in a frequentist framework. Finally, we briefly discuss the relationship between the reach in exclusion and discovery as defined in the analyses of the Tevatron collaborations, and document the approximations used in the text. For the derivation of Eq.~\ref{LLR}--\ref{quadcomb} we follow closely the discussion in Ref.~\cite{Aad:2009wy}. This appendix should not be considered an exhaustive summary of statistical methods used in Higgs searches; rather, our goal is to derive only those formulae relevant for the generation of our plots, and discuss their simplest limits.

Discovery occurs when obtaining the data in the background-only hypothesis is determined to be more unlikely than a one-sided 5$\sigma$ fluctuation of a normal distribution. In the limit of a large number of events and neglecting systematic errors, the background and signal-plus-background distributions for a single random variable can be taken to be approximately normal with (mean, variance) pairs $(B,B)$ and $(S+B,S+B)$, respectively. Then an estimate for the discovery potential is obtained by testing the background-only hypothesis against data given by $N=S+B$. Such data reflects an upward fluctuation of size $S$ above the mean. Normalizing to one standard deviation results in the familiar expression for statistical significance of a counting experiment,
\begin{equation}
n=S/\sqrt{B}.
\end{equation}

An exclusion limit, on the other hand, is an upper bound on the amount of signal that could be present, and occurs for values of signal such that the data in the background-plus-signal hypothesis is found to be less likely than a downward fluctuation of magnitude greater than or equal to some fixed statistical significance $n$ in a Gaussian distribution. In the same Gaussian approximation taken for discovery above, one can test the background-plus-signal hypothesis against data $N=B$. This time, the data is a downward fluctuation of size $S$ below the mean, for a statistical significance of
\begin{equation}
n=S/\sqrt{B+S}.
\end{equation}
Fixing $n$, one can solve for the upper bound on $S$,
\begin{equation}
S=\frac{1}{2}n(\sqrt{4B+n^2}+n),
\end{equation}
In the limit $\sqrt{B}\gg 1$ this reduces to $S=n\sqrt{B}$, so that the estimate $n=S/\sqrt{B}$ can be used either to determine discovery significance or to set exclusion limits on the signal.

The experimental groups at both colliders employ much more accurate calculations of the statistical significance of their results, which do not rely on the Gaussian approximation, combine multiple channels, and include extensive treatments of systematic errors. For brevity, we review primarily the basics of the methods to be used at the LHC. In Refs.~\cite{Aad:2009wy} and~\cite{Ball:2007zza}, the ATLAS and CMS collaborations detail a frequentist analysis which seeks to answer the same question as above, how unlikely is it to obtain data at least as unlikely as the measured set in a fixed model? As in the simple estimate above, it is a hypothesis test based on an expected number of counts $R\times S_{ij}+B_{ij}$, where $i,j$ span the bins and channels and $R$ is a universal rescaling of the expected signal $S_{ij}$ of some reference model, such as the SM or the MSSM. The probability of obtaining data $N_{ij}$ is given by the product of Poisson distributions,
\begin{equation}
P(\vec{N}|R,\vec{S},\vec{B})=\prod_{i,j}\frac{(R\times S_{ij}+B_{ij})^{N_{ij}}e^{-(R\times S_{ij}+B_{ij})}}{N_{ij}!}
\label{prodpos}
\end{equation}
With fixed $N_{ij}$ and variable $R$, this is equivalent to the likelihood function $L(R|\vec{N},\vec{S},\vec{B})$. Note that $S_{ij}$ and $B_{ij}$ implicitly depend on a set of parameters $\theta_k$, which includes more elementary quantities that define the model and account for any systematic errors. For convenience we abbreviate the likelihood function as $L(R|\vec{\theta})$. Now we form the statistic $q(R)$, given by the log-likelihood ratio (LLR)
\begin{equation}
q(R)=-2\log\frac{L(R,\hat{\hat{\theta}})}{L(\hat{R},\hat{\theta})},
\label{LLR}
\end{equation}
where the denominator is evaluated for $(\hat{R},\hat{\theta})$ which maximize $L$, and the numerator is evaluated for $\hat{\hat{\theta}}$ which maximizes $L$ with fixed $R$. If Nature realizes a particular value of $R$ which we label $R'$, and we measure $q(R)$ for $R\neq R'$, we should find that $q(R)$ is distributed such that large values are favored. This real distribution of $q(R)$ is given by some $f(q(R)|R')$. Since we do not know $R'$, we can test the cumulative probability of obtaining a measured value $q_{obs}(R)$ at least as unlikely in a hypothesized universe with $R'=R$ by computing the p-value
\begin{equation}
p=\int_{q_{obs}}^{\infty}f(q(R)|R)dq(R).
\label{pval}
\end{equation}

According to Wilks' theorem~\cite{Wilks}, in the limit of a large data sample (integrated luminosities $\gtrsim 2$ fb$^{-1}$), $f(q(R)|R)$ is a $\chi^2$-distribution with $m$ degrees of freedom, where $m$ is the difference between the number of free parameters in the numerator and denominator of the LLR. Since our only free parameter is $R$, we have $m=1$. However, a small complication arises from restricting $R$ and $\hat{R}$ to nonnegative values, as is done in the experimental analyses. These serve to modify the distribution $f(q(R)|R)$ from $\chi^2(q)$ to
\begin{equation}
f(q(R)|R) = \frac{1}{2}\chi^2(q)+\frac{1}{2}\delta(q),
\label{fdef}
\end{equation}
because (for example, in the $R=0$ hypothesis taken for discovery significance) half the data should show a fluctuation below the background. The $R\geq 0$ restriction then requires $\hat{R}\equiv 0$, producing a pileup at $q=0$. We take Eq.~\ref{fdef} as a definition for both discovery and exclusion and refer the reader to Ref.~\cite{Aad:2009wy} for further details on the $\delta(q)$ modification.

It is then convenient to change variables to $u\equiv\sqrt{q(R)}$, which is distributed as a linear combination of a standard half-normal distribution and a $\delta$-function, again with equal weights. The standard half-normal piece arises because the square of a normally-distributed variable is $\chi^2$-distributed in one degree of freedom, but inverting this relation by taking the square root leaves only positive values of $u$ well-defined. The $\delta$-function piece is present because $u=0$ is just as likely as $q=0$. Therefore we have for the distribution of $u$
\begin{equation}
u \sim \Theta(u)\sqrt{\frac{1}{2\pi}}e^{-(u)^2/2}+\frac{1}{2}\delta(u).
\end{equation}
Since we are interested only in cases where $q_{obs}(R)>0$, Eq.~\ref{pval} can be rewritten in the new variable as
\begin{equation}
p=\sqrt{\frac{1}{2\pi}}\int_{\sqrt{q_{obs}}}^{\infty}e^{-u^2/2}du.
\label{pv}
\end{equation}
According to our definition relating p-values to one-sided Gaussian fluctuations, Eq.~\ref{pv} implies a statistical significance $n=\sqrt{q_{obs}(R)}$.




For multiple independent bins/channels the denominator of the combined likelihood ratio must be maximized over $R$, and this $\hat{R}$ will in general be different from the $\hat{R}_{ij}$ obtained from studying the bins/channels individually. However, in the absence of real data, we can still estimate the expected combined statistical significance of a hypothesis characterized by $R$ from the statistical significances of each separate bin/channel. First, we set the data to the mean values under a true distribution specified by $R'$, $N_{ij}=R'\times S_{ij}+B_{ij}$ (known as the ``Asimov data set"~\cite{Asimov}). The likelihood maxima for the individual bins/channels will all be achieved for $\hat{R}_{ij}\approx{R'}$, and therefore in the combined likelihood ratio $L(\hat{R},\hat{\theta})$ will also be maximized by $\hat{R}\approx R'$. This implies that the combined likelihood ratio factorizes into a product over the likelihood ratios for each bin/channel:
\begin{eqnarray}
\frac{L(R,\hat{\hat{\theta}})}{L(\hat{R},\hat{\theta})}&=&\prod_{ij}\frac{e^{-B_{ij}-R\times S_{ij}}(R\times S_{ij}+B_{ij})^{R'\times S_{ij}+B_{ij}}}{e^{-B_{ij}-R'\times S_{ij}}(R'\times S_{ij}+B_{ij})^{R'\times S_{ij}+B_{ij}}}\nonumber\\
&=&\prod_{ij}\frac{L(R,\hat{\hat{\theta}})}{L(\hat{R}_{ij},\hat{\theta})}.
\end{eqnarray}
(Note that because the channels are independent, the nuisance parameters $\theta_k$ are maximized in the combined ratio by the same values as for the individual channels.) Then we can compute the combined statistical significance:
\begin{eqnarray}
\sqrt{q(R)}&=&\sqrt{-2\sum_{ij}\log\frac{L(R,\hat{\hat{\theta}})}{L(\hat{R}_{ij},\hat{\theta})}}\nonumber\\
&=&\sqrt{\sum_{ij}q_{ij}(R)}.
\label{quadcomb}
\end{eqnarray}
Since $\sqrt{q_{ij}(R)}$ is the statistical significance obtained from one bin in a single channel in the $\chi^2$ limit discussed above, we conclude that the statistical significances derived from individual bins/channels can be added in quadrature to obtain an estimate for the combined significance.

If we assume $B_{ij}$ is large, we can approximate the Poisson distributions in the LLR by normal distributions. Testing the background-only hypothesis $R=0$ in the presence of an Asimov set with $R'=1$, we obtain in the limit of $B_{ij}\gg S_{ij}\gg 1$:
\begin{eqnarray}
q(0)&=&-2\log\frac{\prod_{ij}\frac{1}{\sqrt{2\pi B_{ij}}}\left(e^{\frac{-(S_{ij}+B_{ij}-B_{ij})^2}{2B_{ij}}}\right)}{\prod_{ij}\frac{1}{\sqrt{2\pi( B_{ij}+S_{ij})}}\left(e^{\frac{-(S_{ij}+B_{ij}-S_{ij}-B_{ij})^2}{2(B_{ij}+S_{ij})}}\right)}\nonumber\\
&=&\sum_{ij}S_{ij}^2/B_{ij}-\log(1+S_{ij}/B_{ij})\nonumber\\
&\approx&\sum_{ij}S_{ij}^2/B_{ij}.
\end{eqnarray}
Thus, we can add $S_{ij}/\sqrt{B_{ij}}$ values for each channel in quadrature to approximate the combined statistical significance of a discovery.

For the case of limit-setting in the signal+background hypothesis the procedure is similar, with the Asimov data set taken to be $N_{ij}=B_{ij}$ corresponding to $R'=0$, and $R$ left unfixed. In this case we obtain
\begin{eqnarray}
q(R)&=&\sum_{ij}R^2\times S_{ij}^2/(B_{ij}+R\times S_{ij})+\log(1+R\times S_{ij}/B_{ij})\nonumber\\
&\approx& \sum_{ij}R^2\times S_{ij}^2/B_{ij}.
\label{limset}
\end{eqnarray}
In the second line we have assumed the limit $B_{ij}\gg R\times S_{ij}\gg 1$. Now we would like to fix to a confidence level of $n\sigma$ and solve for $R$. The solution, which we will call $R^{(n)}$, is the expected upper bound on models which differ from the reference model by a universal rescaling of the production cross sections for all channels. Eq.~\ref{limset} becomes
\begin{eqnarray}
n&=&R^{(n)}\sqrt{\sum_{ij}S^2_{ij}/B_{ij}}\nonumber\\
&=&R^{(n)}\sqrt{\sum_{ij}(n/R^{(n)}_{ij})^2},
\label{limset2}
\end{eqnarray}
where we have used $R^{(n)}_{ij}\times S_{ij}=n\sqrt{B_{ij}}$. Therefore, we obtain the combination formula
\begin{equation}
\frac{1}{(R^{(n)})^2}=\sum_{ij}\frac{1}{(R^{(n)}_{ij})^2}.
\label{invquad}
\end{equation}
In other words, the combined upper bound on the signal normalized to the signal in a reference model can be estimated by adding in inverse quadrature the upper bounds on this quantity from each individual bin and channel.

In practice, the Tevatron collaborations compute exclusion limits in a different way from Ref.~\cite{Aad:2009wy}, using both Bayesian and alternate frequentist methods. However, in the limit $B_{ij}\gg S_{ij}\gg 1$, Eq.~\ref{limset2} and~\ref{invquad} are still valid. In the text, we use this inverse quadrature prescription to combine existing upper bounds on the signal from various channels provided by CDF and D0, normalized to the expected signals for those channels in the MSSM. The efficacy of Eq.~\ref{invquad} for the Tevatron Higgs searches was tested in Ref.~\cite{Draper:2009fh} by direct comparison with the combined limits presented by the collaborations. The collaborations' combination was derived using a full likelihood ratio analysis, and the na\"{\i}ve combination was found to give a qualitatively very good match. This implies that $B_{ij}\gg R^{(n)}_{ij}\times S_{ij}\gg 1$ is indeed a good approximation for Higgs searches at the Tevatron, and therefore the $S/\sqrt{B}$ estimate of statistical significance is valid for both discovery and exclusion. Furthermore, we can make projections for the future limit-setting potential of the Tevatron according to the scaling law $R^{(n)}\propto\epsilon^{-1}L^{-1/2}$, where $\epsilon$ is the signal efficiency and $L$ is the integrated luminosity.

For the LHC we use tables of expected $S_{ij}$ and $B_{ij}$ given by ATLAS and CMS in Refs.~\cite{Aad:2009wy} and~\cite{Ball:2007zza} to compute the Poisson distributions and resulting discovery significances in the $\sqrt{q(R)}$ approximation for a given luminosity, combining significances with Eq.~\ref{quadcomb} and avoiding the $S/\sqrt{B}$ approximation which is poor for some LHC channels.

Finally, to obtain a meaningful combination of the LHC and Tevatron reaches, we must convert the Tevatron projected upper limits into potential discovery significances using $S_{ij}/\sqrt{B_{ij}}\approx n/R^{(n)}_{ij}$. Thus far in this appendix we have purposefully avoided referring to ``confidence levels" or explicit values of $n$ used for exclusion, because the Tevatron collaborations define 95\% C.L. differently from $(1-p)$ in the signal+background hypothesis, and therefore it cannot be immediately converted (even in the $B_{ij}\gg S_{ij}\gg 1$ limit) to a statistical significance of discovery in the background-only hypothesis via Eq.~\ref{pv}. Instead, in the $CL_s$ formalism used by the Tevatron groups (reviewed in Ref.~\cite{Amsler:2008zzb}), a 95\% C.L. exclusion reach is effectively identified in this limit with a $1.96\sigma$ discovery reach for the Asimov datasets. With the identification of $n=1.96$ for the Tevatron, we combine the expected statistical significances from both colliders in quadrature.

For completeness, it is worth mentioning a few additional formulas useful for interpreting experimental results from the Tevatron. The collaborations typically present both the expected and observed 95\% C.L. upper limits on $R$. In our study we used only the expected limits, which assume that the data reflects the average number of counts from pure background, so that we could make predictions for the future expected reach of the collider. (Observed limits, on the other hand, may contain statistical fluctuations that will probably be different in the future.) However, if the Higgs is present, then it may generate a real excess in the data and cause the observed limit to be weaker than the expected limit. It is useful to know the relationship between the 95\% C.L. upper bounds $R_{obs}$ and $R_{exp}$ if the data is actually $S_{ij}+B_{ij}$, rather than just $B_{ij}$. Again in the $B_{ij}\gg R^{(n)}_{ij}\times S_{ij}\gg 1$ limit, it can be shown that
\begin{equation}
R_{obs}\approx R_{exp}+1,
\end{equation}
where this holds for the limits derived from individual bins and channels as well as for the combined limit.
On the other hand, even if the signal is present, due to statistical fluctuations the data may not be exactly $S_{ij}+B_{ij}$. Therefore, given only $R_{obs}$ and $R_{exp}$, it is also useful to have an approximate formula for the observed statistical significance of discovery. Extracting the approximate data value from $R_{obs}$ in the signal + background hypothesis and computing the significance $n_{obs}$ in the background-only hypothesis, we find
\begin{equation}
n_{obs}\approx 2\left(\frac{R_{obs}}{R_{exp}}-1\right).
\end{equation}
Let us stress that this formula only holds for the $R$ values derived from individual channels or combinations where a single channel dominates, not for the combinations with multiple significant channels. For the latter case more information is required to extract the observed discovery significance.

\newpage

\setcounter{equation}{0}

\section{Tevatron+LHC Combinations for $CP$-conserving Benchmark Scenarios}

\renewcommand{\theequation}{\Alph{section}.\arabic{equation}}

For completeness, we present here the estimated discovery significance achievable in two standard $CP$-conserving benchmark scenarios from the combination of SM-like Higgs search data from the Tevatron and the low-luminosity LHC.
The Maximal Mixing scenario is defined by
\begin{eqnarray*}
M_S=1 \mbox{ TeV, }& &a_t=\sqrt{6}M_S\mbox{,}\nonumber\\
\mu=200 \mbox{ GeV, }& &M_2=200\mbox{ GeV,}\nonumber\\
A_b=A_t\mbox{, }& &m_{\tilde{g}}=0.8M_S
\end{eqnarray*}
where $a_t$ is the stop squark mixing parameter given by $a_t\equiv A_t-\mu/\tan\beta$. In the decoupling limit, this choice of parameters saturates the upper bound on the SM-like Higgs mass. The Minimal Mixing scenario is defined by
\begin{eqnarray*}
M_S=2 \mbox{ TeV, }& &a_t=0\mbox{,}\nonumber\\
\mu=200 \mbox{ GeV, }& &M_2=200\mbox{ GeV,}\nonumber\\
A_b=A_t\mbox{, }& &m_{\tilde{g}}=0.8M_S
\end{eqnarray*}
and produces an SM-like Higgs with a mass just above the LEP bound in most of the parameter space. A detailed discussion of the prospects for both of these scenarios at the Tevatron is given in Ref.~\cite{Draper:2009fh}.

\begin{figure}[!htbp]
\begin{center}
\begin{tabular}{cc}
\includegraphics[width=0.40\textwidth]{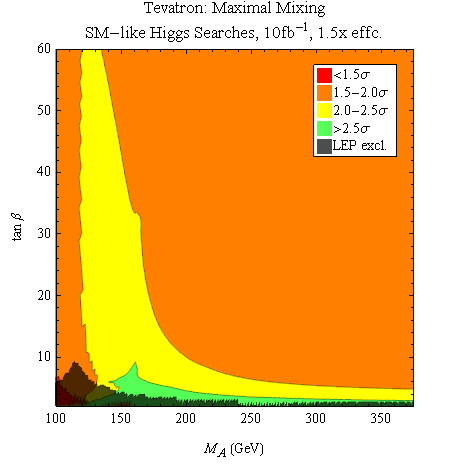} &
\includegraphics[width=0.40\textwidth]{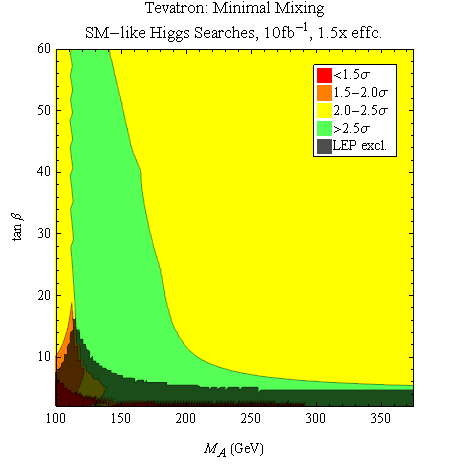} \\
\includegraphics[width=0.40\textwidth]{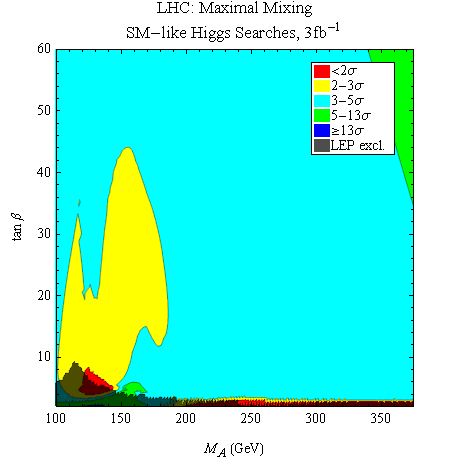} &
\includegraphics[width=0.40\textwidth]{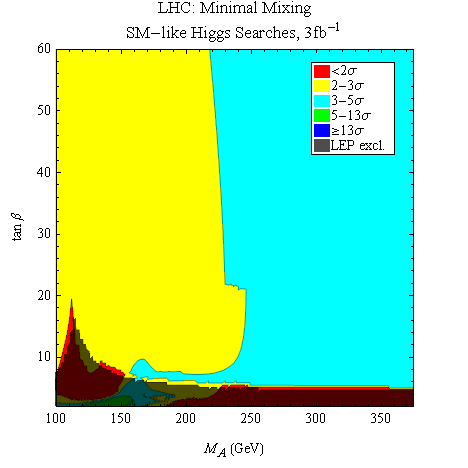} \\
\includegraphics[width=0.40\textwidth]{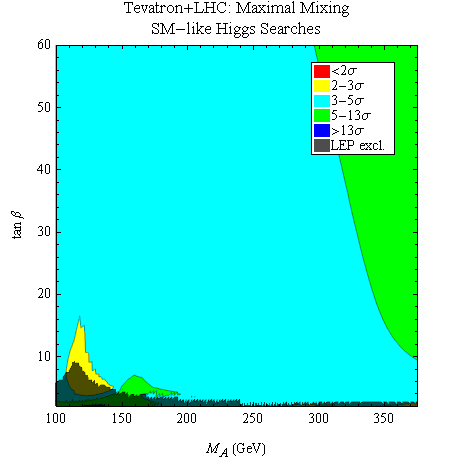} &
\includegraphics[width=0.40\textwidth]{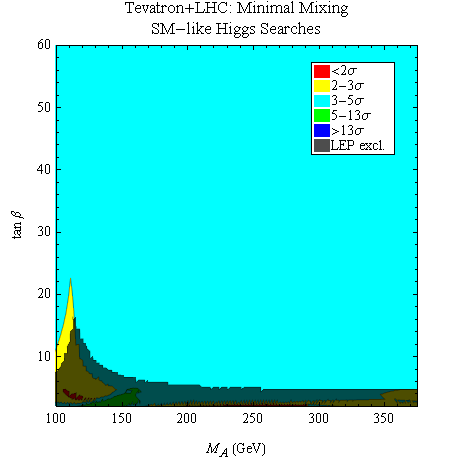} \\
\end{tabular}
\caption{Tevatron and LHC projected discovery significances for the $CP$-conserving Maximal Mixing scenario (\textit{left}) and the Minimal Mixing scenario (\textit{right}). Row 1 gives results for the Tevatron, row 2 for the LHC, and row 3 gives the combination. Note that the contours for the Tevatron differ from those used for the LHC and the combination.}
\label{cpconstitans}
\end{center}
\end{figure}

Because the value of $\mu$ is small in both of these scenarios, the branching ratios of the Higgs states to $\tau^+\tau^-$ never receive significant enhancement. Consequently, the $\gamma\gamma$ channel becomes the strongest search mode at the LHC for SM-like Higgs states. This channel prefers heavier Higgs masses, which suggests another instance of complementarity with the Tevatron, where lighter Higgs masses strengthen the $b\bar{b}$ channel reach. In Fig.~\ref{cpconstitans} we give estimated discovery reach of the Tevatron, the LHC, and the combination for each of these scenarios. Note that the significance contours used for the Tevatron differ from those used for the LHC and the combination, but that the Tevatron and LHC searches are strongest in complementary regions. The primary conclusion is that in concert with the Tevatron data, the low-luminosity LHC can probe the $CP$-conserving MSSM at greater than $3\sigma$.

\end{document}